\documentclass[
]{ceurart}

\usepackage{listings}
\begin{document}

\copyrightyear{2024}
\copyrightclause{Copyright for this paper by its authors.
  Use permitted under Creative Commons License Attribution 4.0
  International (CC BY 4.0).}

\conference{eCom’24: ACM SIGIR Workshop on eCommerce, July 18, 2024, Washington, DC, USA}

\title{Long or Short or Both? An Exploration on Lookback Time Windows of Behavioral Features in Product Search Ranking}


\author{Qi Liu}[%
email=qi.liu@walmart.com,
]

\author{Atul Singh}[%
email=atul.singh@walmart.com,
]

\author{Jingbo Liu}[%
email=jingbo.liu@walmart.com,
]

\author{Cun Mu}[%
email=cun.mu@walmart.com,
]

\author{Zheng Yan}[%
email=zheng.yan0@walmart.com,
]

\author{Jan Pedersen}[%
email=jan.pedersen@walmart.com,
]
\address{Walmart Global Tech, Hoboken, NJ, U.S.A.}


\begin{abstract}
Customer shopping behavioral features are core to product search ranking models in eCommerce. In this paper, we investigate the effect of lookback time windows when aggregating these features at the (query, product) level over history. By studying the pros and cons of using long and short time windows, we propose a novel approach to integrating these historical behavioral features of different time windows. In particular, we address the criticality of using query-level vertical signals in ranking models to effectively aggregate all information from different behavioral features. Anecdotal evidence for the proposed approach is also provided using live product search traffic on \textit{Walmart.com}.


\end{abstract}

\begin{keywords}
  Online shopping \sep 
  product search ranking \sep
  learning to rank \sep
  feature engineering \sep
  behavioral features 
\end{keywords}

\maketitle

\section{Introduction}
\label{sec:intro}

Online shopping has become an indispensable part of people's daily lives due to its convenience, wide selection, cost-effectiveness, and mobile accessibility. With an ever increasing catalog size, product search ranking system \cite{sorokina2016amazon, trotman2017architecture, brenner2018end, tsagkias2021challenges, eletreby2022machine, wu2022multi} has been playing a pivotal role in serving customers by ranking relevant products at the top of their search results. 


At the heart of every modern eCommerce product search ranking system lies a machine-learned ranking model. For example, LambdaRank/MART \cite{burges2006learning, burges2010ranknet} leverages gradient boosting machines \cite{friedman2001greedy}, and neural ranker \cite{guo2020deep} employs deep learning techniques. These models evaluate and assign scores to each product based on a wide range of input signals derived from diverse sources, including user behaviors, query intents, product attributes, seller reputations, and sophisticated interactions among them.

Out of these many hundreds or even thousands of signals, behavioral features hold significant importance as they are generated through direct interactions between customers and products, encompassing actions like impressions, clicks, add-to-carts (ATCs), purchases, and others. Several studies \cite{sorokina2016amazon, Chapelle2011, gupta2020treating, han2022addressing, Hendriksen2020} have emphasized the pivotal role of such implicit relevance feedback \cite{rocchio1971relevance} in product ranking. In the eCommerce context, customers are the ultimate authorities in determining the relevance of products for a given query, particularly when their judgment is backed by their purchasing decisions. Moreover, such logged customer feedback is abundant and cheap to obtain in nowadays operational systems. Hence, it is very natural that (query, product)-level behavioral features are the ones that ranking models rely on the most when ranking products.

In spite of the rich and growing literature on leveraging (query, product)-level behavioral features and their variants in product search ranking, one much unaddressed problem is \textit{what the lookback time window should be used to aggregate the customer engagement at the (query, product) level}. This is a very critical design question for all practitioners when applying these essential behavioral features in their ranking systems. In this paper, we will share our empirical insights from our first-hand industrial experience. In particular, we explore behavioral features aggregated over different lookback time windows and study their respective effects on product search ranking. Based upon the pros and cons of using long and short time windows, we propose a principled approach to integrating both sets of behavioral features into the model. The effectiveness of this hybrid model is justified on real product search traffic at \textit{Walmart.com} through online A/B tests.

The remainder of the paper is organized as follows. 
Section~\ref{sec:CustomerEngagementSignals} discusses the pros and cons of using behavioral features with long and/or short windows. Section~\ref{sec:Both} details the proposed enhancement to achieve the best integration of both long and short windows. Section~\ref{sec:Experiment} describes the comprehensive online A/B experiment conducted to evaluate the proposed ranking model. Finally, Section~\ref{sec:conclusion} summarizes our findings and draws conclusions.

\section{Long or Short or Both?}\label{sec:CustomerEngagementSignals}

In this section, we will explore the effect of different lookback time window lengths when leveraging behavioral features in product search ranking models. Three types of (query, product)-level user engagement are considered: click rate, add-to-cart (ATC) rate, and order rate. To compute these rates, for a given query ($q$) - product ($p$) pair ($q, p$), we employ the Beta-Binomial Bayesian model and derive behavioral feature values $br_{q, p}$ as the posterior mean of the following Beta distribution,

\vspace{-6mm}



\begin{equation}
     \textbf{Beta} \Bigg(\sum_{t \in T} b_{q,p}^{(t)} + \alpha, \sum_{t \in T} e_{q,p}^{(t)} - \sum_{t \in T} b_{q,p}^{(t)} + \beta \Bigg),
\end{equation}
where $\alpha$ and $\beta$ specify the prior Beta distribution, $b_{q,p}^{(t)}$ is the raw count of the behavior (clicks, ATCs, or orders) frequency for ($q, p$) on day $t$, $e_{q,p}^{(t)}$ is the raw count of customer examines for ($q, p$) on day $t$, and $T$ is the collection of lookback dates we use to aggregate the engagement data. In particular, the following behavioral features are output to our ranking model,
\begin{equation}\label{eqn:bf}
     br_{q, p} = \frac{\sum_{t \in T} b_{q,p}^{(t)} + \alpha}{\sum_{t \in T} e_{q,p}^{(t)} + \alpha + \beta},
\end{equation}
which is quite similar to the behavioral features defined in \cite{karmaker2017application} but smoothed with prior in order to better address the cold start problem \cite{han2022addressing}. 

As shown in equation \ref{eqn:bf}, one critical factor influencing the values and interpretation of behavioral features is the lookback time window length $|T|$ used to aggregate engagements. Utilizing a longer time window captures long-term customer engagement patterns but may overlook recent trends. Conversely, a short time window highlights more short-term behaviors but may not accurately reflect enduring customer interests. Both long and short time windows for aggregating behavioral features present distinct advantages and disadvantages outlined in Table~\ref{tbl:pros_cons}.

To investigate the impacts of long- and short-term behavioral features, we define 2 years ($|T| = 730$) as the long lookback time window and 1 month ($|T| = 30$) as the short one, and we specify the ranking model with only 2-year behavioral features as the baseline. Three distinct ranking models with different designs on the window lengths are proposed below.

\vspace{-2mm}
\begin{itemize}
    \item \textbf{Baseline Model:} model with only 2-year behavioral features.
    \item \textbf{Model A:} model with only 1-month behavioral features.
    \item \textbf{Model B:} model with both 2-year and 1-month behavioral features.
\end{itemize}
\vspace{-2mm}

\begin{table}
\centering
\begin{tabular}{lll}
\toprule
\begin{tabular}[c]{@{}l@{}}\textbf{Window} \\ \textbf{Lengths}\end{tabular} & \multicolumn{1}{c}{\textbf{Long}}                                                    & \multicolumn{1}{c}{\textbf{Short}}                                               \\ 
\midrule
\textbf{Pros}                                                               & \begin{tabular}[c]{@{}l@{}}• rich in historical engagement data\\ • robust to noise\end{tabular}           & \begin{tabular}[c]{@{}l@{}}• good at capturing recent behavioral \\ changes from customers \\ • friendly to new products \end{tabular}   \\ 
\midrule
\textbf{Cons}                                                               & \begin{tabular}[c]{@{}l@{}}• insensitive to recent behavioral \\ changes from customers \\ • frictional to new products \end{tabular}  & \begin{tabular}[c]{@{}l@{}}• sparse in historical engagement data \\ • prone to noise\end{tabular} \\ 
\bottomrule
\end{tabular}
\caption{\textbf{Pros and cons of long and short time windows for behavioral features.} The longer the time window, the more historical engagement observations the features capture, which leads to less sparsity and higher coverage in model training and inference. The shorter the time window, the more it captures recent online shopping trends due to customer behavioral changes, new product launches, etc.}
\vspace{-5mm}
\label{tbl:pros_cons}
\end{table}

Our search ranking models are trained using XGBoost \cite{chen2016xgboost} with the Learning-to-Rank (LTR) framework \cite{liu2009} very similar to \cite{karmaker2017application} by utilizing data from a truncated historical period of online customer search traffic on \textit{Walmart.com} for model training.

To explore the best usage of lookback time windows, we conducted multiple interleaving tests \cite{chapelle2012large}, each comparing one proposed model against the baseline model. These tests were performed on a substantial volume of online customer traffic to compare their reactions to different ranking models. Specifically, for each test, we compare the percentage of searches that result in customer engagements between the Control and Variant groups using their respective ranking models. The results are further segmented by different verticals—specific business niches tailored to particular shopping needs. We currently categorize our search queries into six verticals: Food, Consumables, Home, Hardlines, Fashion, and ETS (Electronics, Toys, and Seasonal), with the latter four collectively categorized as General Merchandise (GM).

\subsection{Only Long / Short Window}\label{subsec:test1}

The first interleaving test is configured as follows: 
\begin{itemize}
    \item \textbf{Control:} Baseline Model (2-year behavioral features only),
    \item \textbf{Variant:} Model A (1-month behavioral features only),
\end{itemize}
with the purpose to separately examine and compare the individual impact of 2-year and 1-month behavioral features on search ranking models.

The test result is presented in Table~\ref{tbl:test1}. Although Model A demonstrates an overall insignificant decline compared to the baseline, when zooming into each business vertical, we find very interesting stories. Model A exhibits a significant decline in Food and a trending decline in Consumables. Conversely, it demonstrates a significant lift in the ETS and, more generally, positive changes across most General Merchandise verticals. This corroborates that short-term behavioral features are more informative in an environment that is more dynamic in terms of both inventory assortment and customers' shopping behaviors. In contrast, long-term features are more advantageous for business units Food and Consumables, which typically display more stable and enduring shopping patterns. Therefore, it is very tempting to employ both types of features in the ranking model to leverage their combined strengths. Similar ideas of combining session and historical customer search behaviors per each customer are also investigated in web search personalization \cite{Bennett2012}, but to the best of knowledge, our work is the first one to explore combining (query, product)-level historical behavior features over different lookback time windows in product search ranking.


\begin{table}
\centering
\begin{tabular}{lccccccc}
\toprule
\textbf{Vertical} & Food    & Consumables & Home    & ETS     & Hardlines & Fashion & {Overall} \\ 
\midrule
\textbf{Change}   & $-0.63\%^{*}$ & $-0.67\%$     & $-0.34\%$ & $+3.79\%^{*}$ & $+1.51\%$   & $+1.06\%$ & ${-0.28\%}$ \\
\bottomrule
\end{tabular}
\caption{\textbf{Result of \% searches with engagement for Test 1.} The control is the baseline model (only using 2-year behavioral features), and the variant is Model A (only using 1-month behavioral features). The significance level of all tests throughout this paper is set to be 0.1, and the statistically significant results are starred in the tables.}
\vspace{-3mm}
\label{tbl:test1}
\end{table}

\subsection{Both Long \& Short Windows}\label{subsec:test2}
With the insight from the previous subsection, we set up the second interleaving test as follows: 
\begin{itemize}
    \item \textbf{Control:} Baseline Model (2-year behavioral features only),
    \item \textbf{Variant:} Model B (2-year and 1-month behavioral features),
\end{itemize}
with the purpose to examine the combined impact of using both 2-year and 1-month behavioral features in ranking.

The test result is presented in Table~\ref{tbl:test2}. To our surprise, Model B performs quite sub-optimally overall with the degradation in Food, Consumables, and ETS verticals. This suggests that combining both long- and short-term behavioral features in this vanilla manner not only fails to provide gains in ranking performance but also leads to further declines. One possible reason for the negativity is the lack of flexibility in our ranking model to leverage different behavioral features accordingly. For instance, the Food vertical should ideally leverage the 2-year behavioral features as extensively as possible. However, adding 1-month features dilutes the positive effect of the 2-year features, negatively interfering with the overall contribution of behavioral features. Conversely, in verticals such as ETS and Hardlines, where 1-month features are more advantageous, the inclusion of 2-year features can similarly impair performance. 


\begin{table}
\centering
\begin{tabular}{lccccccc}
\toprule
\textbf{Vertical} & Food    & Consumables & Home    & ETS     & Hardlines & Fashion & {Overall} \\ 
\midrule
\textbf{Change}   & $-0.46\%^{*}$ & $-0.51\%^{*}$     & $+0.29\%$ & $-0.46\%$ & $-1.07\%$   & $+1.72\%$ & ${-0.41\%}^{*}$\\
\bottomrule
\end{tabular}
\caption{\textbf{Result of \% searches with engagement for Test 2.} The control is the baseline model (only using 2-year behavioral features), and the variant is Model B (using both 1-month and 2-year behavioral features).}
\vspace{-3mm}
\label{tbl:test2}
\end{table}

\section{How to Integrate Both?}\label{sec:Both}

Different verticals exhibit different patterns of trending effects in customer behaviors. For instance, Fashion such as ``clothes'' may be significantly influenced by recent trends affecting their popularity and customer interactions. In contrast, Food and Consumables such as ``milk'' and ``toilet paper'' tend to show more stability over time and are predominantly shaped by long-term engagement patterns.

Based on observations from tests in Sections \ref{subsec:test1} and \ref{subsec:test2}, to improve the model performance with combined behavioral features of both long and short windows, we consider making our ranking model more query context-aware by incorporating one-hot encoded query vertical signals (predicted from the upstream query understanding model) into the model. These query-level vertical signals would better guide our ranking model to leverage behavioral features of different time windows according to different queries. Thus, we propose the fourth ranking model below.
\begin{itemize}
    \item \textbf{Model C:} model with both 2-year and 1-month features, and query-level vertical features.
\end{itemize}

\subsection{Both Long \& Short Windows with Verticals}\label{subsec:test3}

The third interleaving test is configured as follows. 
\begin{itemize}
    \item \textbf{Control:} Baseline Model (2-year behavioral features only),
    \item \textbf{Variant:} Model C (2-year and 1-month behavioral features with the vertical features),
\end{itemize}
with the purpose to examine whether adding query-level vertical features helps better integrate 2-year and 1-month behavioral features in ranking.

The test result, detailed in Table~\ref{tbl:test3}, shows that guided by vertical information, behavioral features are more effectively utilized by the ranking model, leading to significant uplifts across all General Merchandise verticals while rectifying the previous degradation in the Food and Consumables. Model C also shows an overall significant increase of 0.22\% in customer engagement and proves to be the best candidate ranking model among all tested. This demonstrates that incorporating vertical features can indeed enhance the integration of multi-window behavioral features, allowing each to play to its strengths and mitigate its weaknesses.

\begin{table}
\centering
\begin{tabular}{lccccccc}
\toprule
\textbf{Vertical} & Food    & Consumables & Home    & ETS     & Hardlines & Fashion & \textbf{Overall} \\ 
\midrule
\textbf{Change}   & $-0.01\%$ & $+0.02\%$     & $+0.40\%^{*}$ & $+0.78\%^{*}$ & $+1.58\%^{*}$   & $+0.73\%^{*}$ & ${+0.22\%}^{*}$ \\
\bottomrule
\end{tabular}
\caption{\textbf{Result of \% searches with engagement for Test 3.} The control is the baseline model (only using 2-year behavioral features), and the variant is Model C (using both 1-month and 2-year behavioral features along with the vertical features).}
\vspace{-2.5mm}
\label{tbl:test3}
\end{table}

\subsection{Why Does Long \& Short \& Verticals Work?}\label{subsec:test3rca}
The guiding effect that vertical information has on the ranking model in using different behavioral features can also be validated in the model structure. Our ranking model inherently employs a tree structure, where adjacent tree nodes tend to be functionally related. More precisely, if the nodes corresponding to one feature frequently precede those of another specific feature, it suggests that the former, i.e., the upper-level feature, exerts a certain degree of influence/control over the latter, i.e., the lower-level feature, determining when it will activate to impact the model's predictions.

\begin{figure}
\centering
\includegraphics[scale=0.24]{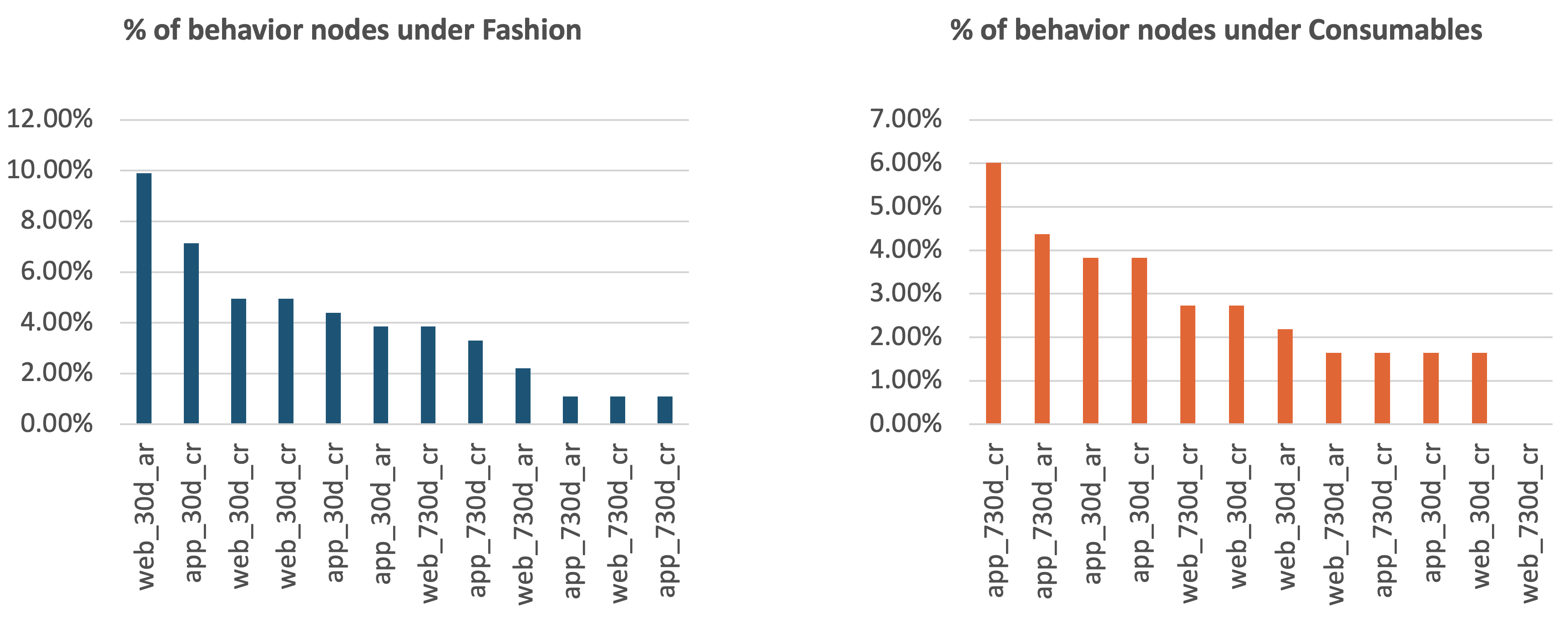}
\caption{\textbf{Distribution of different behavioral nodes under vertical nodes for Fashion and Consumables.} We summarize the percentage of behavioral tree nodes under the verticals. The behavioral feature names have 3 segmented sub-strings: the first indicates the behavioral data source (web and app); the second indicates the time window lengths (730-day and 30-day); the third indicates the behavioral types (cr = click rate; ar = ATC rate; or = order rate).}
\label{fig:node_perc}
\end{figure}

Across all splitting nodes from the trees in Model C, we summarize the distribution of different behavioral nodes under the vertical nodes in Figure~\ref{fig:node_perc}, taking Fashion and Consumables as examples. The results show that 1-month behavioral features are more influential for Fashion queries since they more prevalently occupy the place of the immediate lower level when the current vertical node is Fashion, whereas 2-year behavioral features are more prevalent for Consumables queries. This observation is aligned with our interpretation of the test result in Section~\ref{subsec:test3} given the characteristics of different verticals, and it evinces that introducing query-level vertical signals can help guide our ranking model to better ensemble long- and short-term behavioral features in the sense that different behavioral features can contribute accordingly with respect to different search queries.

\section{A/B Test}\label{sec:Experiment}
After the series of interleaving tests in Sections \ref{sec:CustomerEngagementSignals} and \ref{sec:Both}, we decided to move forward to A/B test with the most promising candidate, Model C, which incorporates both long- and short-term behavioral features along with the query-level vertical signals. Specifically, we conducted a comprehensive A/B test on \textit{Walmart.com} for two weeks to compare Model C against the baseline production model. 

The result, detailed in Table~\ref{tbl:ab_test}, highlights substantial improvements in key search related metrics. This A/B test observation confirms our hypothesis that a vertical-aware ranking model incorporating a hybrid of behavioral features across both long and short time windows can enhance the customer experience for a diverse range of online shopping needs. In addition, the positivity in marketplace GMV clearly indicates that we are also able to better address cold-start problems when introducing short-term behavioral features into the system. 

\begin{table}
\centering
\begin{tabular}{lccccc}
\toprule
\textbf{Metric}   & \begin{tabular}[c]{@{}c@{}}Overall\\ GMV\end{tabular}    & \begin{tabular}[c]{@{}c@{}}Marketplace\\ GMV\end{tabular} & ATC@10    & \begin{tabular}[c]{@{}c@{}}Sessions\\ with ATC\end{tabular} & \begin{tabular}[c]{@{}c@{}}Session\\ Abandonment\end{tabular} \\ 
\midrule
\textbf{Lift}   & $+0.12\%$ & $+0.64\%^{*}$     & $+0.21\%^{*}$ & $+0.22\%^{*}$  & $-0.16\%^{*}$ \\
\bottomrule
\end{tabular}
\caption{\textbf{Online A/B test result of Model C vs. baseline.} These metrics are all calculated at visitor level: 1) \textbf{Overall GMV}: the total Gross Merchandise Value from all kinds of products sold; 2) \textbf{Marketplace GMV}: the Gross Merchandise Value yielded from marketplace products; 3) \textbf{ATC@10}: the percentage of search ATCs coming from top 10 products in search results; 4) \textbf{Sessions with ATC}: percentage search sessions with at least one ATC; 5) \textbf{Session Abandonment Rate}: percentage search sessions without any user engagement (and thus the smaller the better).}
\vspace{-5mm}
\label{tbl:ab_test}
\end{table}

\begin{figure}
\centering
\includegraphics[scale=0.26]{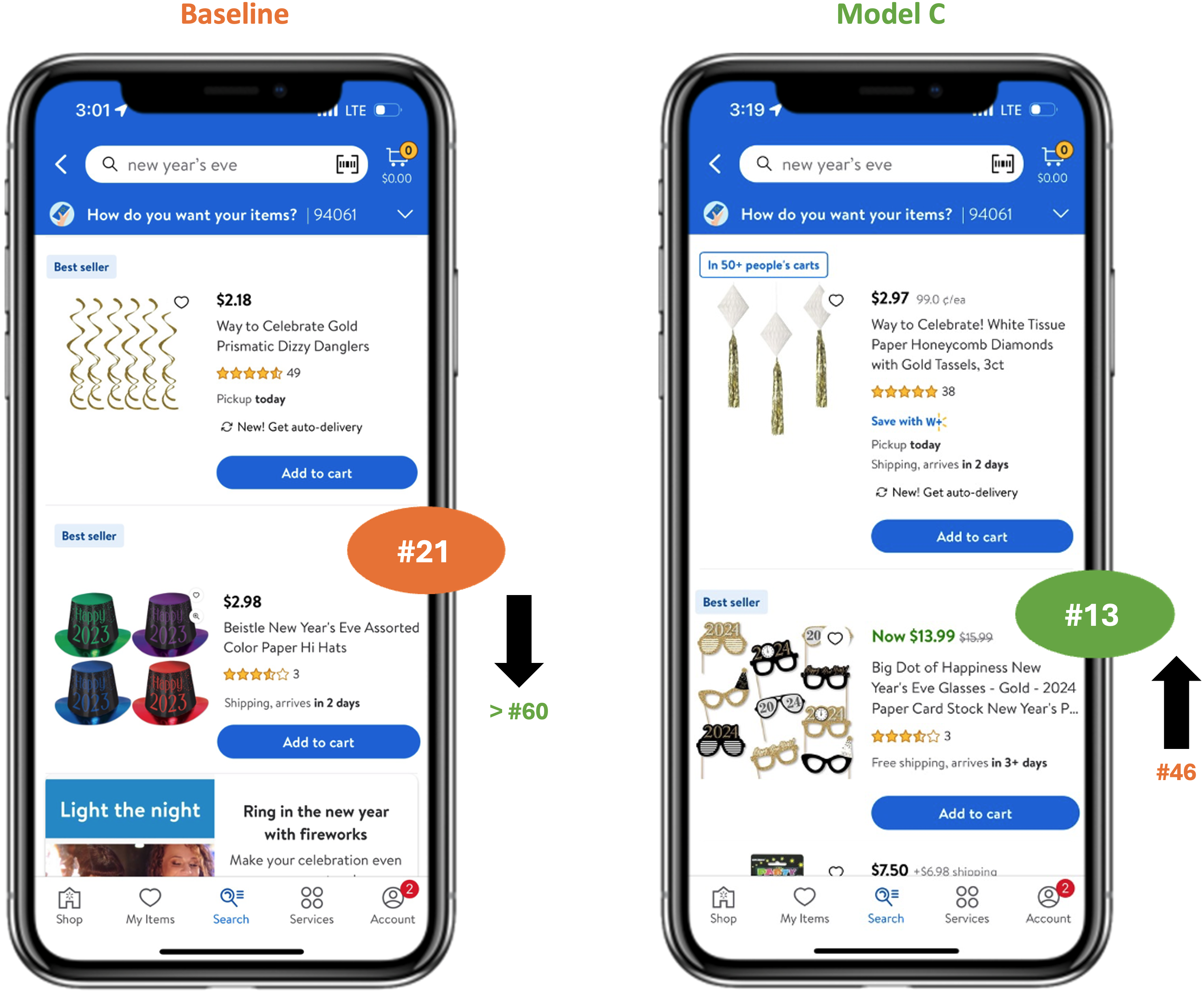}
\caption{\textbf{Example of customer search ranking experience comparing baseline vs. Model C.} This comparison was performed in December 2023 for query \textit{new year's eve}. The baseline incorrectly prioritizes many 2023 New Year's Eve supplies over 2024 products, which should have been ranked higher as the event for 2024 approaches. In contrast, Model C effectively ranks 2024 products higher than those for 2023. In this example, the 2024 glasses item is elevated from position 46 to 13, while the 2023 hat item is demoted from position 21 to beyond 60, moving it off the first page of search results.}
\label{fig:abexample}
\end{figure}

We also present a qualitative example in Figure~\ref{fig:abexample} illustrating the comparison of Model C versus the baseline in terms of user experience from the search ranking. It is clearly demonstrated that utilizing behavioral features from both long and short time windows, along with vertical information, results in a ranking model that prioritizes products with high recent popularity, especially in the General Merchandise categories. This approach ensures that customers are provided with options that are more closely aligned with their current shopping needs. 

\section{Conclusion}
\label{sec:conclusion}

In this paper, we propose a novel product search ranking model that incorporates a hybrid of behavioral features over both long and short lookback time windows with vertical-specific insights. The multi-window design aims to capture customer engagement patterns over varying durations, and the vertical features are purposed to tailor behavioral features more effectively to different online shopping contexts. This approach allows long-term behavioral features to reflect enduring patterns, supporting routine customer journeys, while short-term features capture immediate, trending patterns to enhance discovery customer experiences.

Through comprehensive online testing, we demonstrate that the proposed model significantly outperforms the baseline, which solely utilizes singular time-window behavioral features, by achieving substantial improvements in key evaluation metrics across various verticals, catering to distinct online shopping needs. As a result, the integration of multi-window behavioral features and search context awareness adeptly navigates the complex dynamics of different shopping categories, thereby enhancing customer engagement across all verticals. Consequently, the proposed model not only fulfills the diverse needs of contemporary eCommerce online shopping but also lays a scalable foundation for future enhancements in search ranking systems.

For future work, we intend to expand the feature scope of the search ranking model by incorporating behavioral features from additional time windows, such as 1 week and 1 year. This extension will enable the model to capture a broader spectrum of trending effects, further enhancing its predictive accuracy. Additionally, we plan to introduce more granular query-level signals--e.g., categorical signals, product type signals, etc.--to allow for more nuanced guidance of behavioral features, improving ranking's contextualized capability and enriching the online shopping experience for customers.

\bibliography{main}

\begin{thebibliography}{20}
\expandafter\ifx\csname natexlab\endcsname\relax\def\natexlab#1{#1}\fi
\providecommand{\url}[1]{\texttt{#1}}
\providecommand{\href}[2]{#2}
\providecommand{\path}[1]{#1}
\providecommand{\DOIprefix}{doi:}
\providecommand{\ArXivprefix}{arXiv:}
\providecommand{\URLprefix}{URL: }
\providecommand{\Pubmedprefix}{pmid:}
\providecommand{\doi}[1]{\href{http://dx.doi.org/#1}{\path{#1}}}
\providecommand{\Pubmed}[1]{\href{pmid:#1}{\path{#1}}}
\providecommand{\bibinfo}[2]{#2}
\ifx\xfnm\relax \def\xfnm[#1]{\unskip,\space#1}\fi
\bibitem[{Sorokina and Cantu-Paz(2016)}]{sorokina2016amazon}
\bibinfo{author}{D.~Sorokina}, \bibinfo{author}{E.~Cantu-Paz},
\newblock \bibinfo{title}{Amazon search: The joy of ranking products},
\newblock in: \bibinfo{booktitle}{SIGIR}, \bibinfo{year}{2016}, pp. \bibinfo{pages}{459--460}.
\bibitem[{Trotman et~al.(2017)Trotman, Degenhardt, and Kallumadi}]{trotman2017architecture}
\bibinfo{author}{A.~Trotman}, \bibinfo{author}{J.~Degenhardt}, \bibinfo{author}{S.~Kallumadi},
\newblock \bibinfo{title}{The architecture of ebay search},
\newblock in: \bibinfo{booktitle}{SIGIR eCom}, volume \bibinfo{volume}{2311}, \bibinfo{year}{2017}.
\bibitem[{Brenner et~al.(2018)Brenner, Zhao, Kutiyanawala, and Yan}]{brenner2018end}
\bibinfo{author}{E.~P. Brenner}, \bibinfo{author}{J.~Zhao}, \bibinfo{author}{A.~Kutiyanawala}, \bibinfo{author}{Z.~Yan},
\newblock \bibinfo{title}{End-to-end neural ranking for ecommerce product search},
\newblock in: \bibinfo{booktitle}{SIGIR eCom}, \bibinfo{year}{2018}.
\bibitem[{Tsagkias et~al.(2021)Tsagkias, King, Kallumadi, Murdock, and Rijke}]{tsagkias2021challenges}
\bibinfo{author}{M.~Tsagkias}, \bibinfo{author}{T.~H. King}, \bibinfo{author}{S.~Kallumadi}, \bibinfo{author}{V.~Murdock}, \bibinfo{author}{M.~Rijke},
\newblock \bibinfo{title}{Challenges and research opportunities in ecommerce search and recommendations},
\newblock in: \bibinfo{booktitle}{ACM SIGIR Forum}, \bibinfo{year}{2021}.
\bibitem[{Eletreby et~al.(2022)Eletreby, Mu, Wang, and Mukherjee}]{eletreby2022machine}
\bibinfo{author}{R.~Eletreby}, \bibinfo{author}{C.~Mu}, \bibinfo{author}{Z.~Wang}, \bibinfo{author}{R.~Mukherjee}, \bibinfo{title}{Machine learning based methods and apparatus for automatically generating item rankings}, \bibinfo{year}{2022}. \bibinfo{note}{US Patent App. 17/246,179}.
\bibitem[{Wu et~al.(2022)Wu, Magnani, Chaidaroon, Puthenputhussery, Liao, and Fang}]{wu2022multi}
\bibinfo{author}{X.~Wu}, \bibinfo{author}{A.~Magnani}, \bibinfo{author}{S.~Chaidaroon}, \bibinfo{author}{A.~Puthenputhussery}, \bibinfo{author}{C.~Liao}, \bibinfo{author}{Y.~Fang},
\newblock \bibinfo{title}{A multi-task learning framework for product ranking with bert},
\newblock in: \bibinfo{booktitle}{WWW}, \bibinfo{year}{2022}, pp. \bibinfo{pages}{493--501}.
\bibitem[{Burges et~al.(2006)Burges, Ragno, and Le}]{burges2006learning}
\bibinfo{author}{C.~Burges}, \bibinfo{author}{R.~Ragno}, \bibinfo{author}{Q.~Le},
\newblock \bibinfo{title}{Learning to rank with nonsmooth cost functions},
\newblock \bibinfo{journal}{NeurIPS}  (\bibinfo{year}{2006}).
\bibitem[{Burges(2010)}]{burges2010ranknet}
\bibinfo{author}{C.~Burges},
\newblock \bibinfo{title}{From ranknet to lambdarank to lambdamart: an overview},
\newblock \bibinfo{journal}{Learning} \bibinfo{volume}{11} (\bibinfo{year}{2010}) \bibinfo{pages}{81}.
\bibitem[{Friedman(2001)}]{friedman2001greedy}
\bibinfo{author}{J.~H. Friedman},
\newblock \bibinfo{title}{Greedy function approximation: a gradient boosting machine},
\newblock \bibinfo{journal}{Annals of statistics}  (\bibinfo{year}{2001}) \bibinfo{pages}{1189--1232}.
\bibitem[{Guo et~al.(2020)Guo, Fan, Pang, Yang, Ai, Zamani, Wu, Croft, and Cheng}]{guo2020deep}
\bibinfo{author}{J.~Guo}, \bibinfo{author}{Y.~Fan}, \bibinfo{author}{L.~Pang}, \bibinfo{author}{L.~Yang}, \bibinfo{author}{Q.~Ai}, \bibinfo{author}{H.~Zamani}, \bibinfo{author}{C.~Wu}, \bibinfo{author}{W.~B. Croft}, \bibinfo{author}{X.~Cheng},
\newblock \bibinfo{title}{A deep look into neural ranking models for information retrieval},
\newblock \bibinfo{journal}{Information Processing \& Management} \bibinfo{volume}{57} (\bibinfo{year}{2020}) \bibinfo{pages}{102067}.
\bibitem[{Chapelle and Chang(2011)}]{Chapelle2011}
\bibinfo{author}{O.~Chapelle}, \bibinfo{author}{Y.~Chang},
\newblock \bibinfo{title}{Yahoo! learning to rank challenge overview},
\newblock in: \bibinfo{booktitle}{PMLR}, \bibinfo{year}{2011}, pp. \bibinfo{pages}{1--24}.
\bibitem[{Gupta et~al.(2020)Gupta, Dreossi, Bakus, Lin, and Salaka}]{gupta2020treating}
\bibinfo{author}{P.~Gupta}, \bibinfo{author}{T.~Dreossi}, \bibinfo{author}{J.~Bakus}, \bibinfo{author}{Y.~Lin}, \bibinfo{author}{V.~Salaka},
\newblock \bibinfo{title}{Treating cold start in product search by priors},
\newblock in: \bibinfo{booktitle}{WWW Companion}, \bibinfo{year}{2020}.
\bibitem[{Han et~al.(2022)Han, Castells, Gupta, Xu, and Salaka}]{han2022addressing}
\bibinfo{author}{C.~Han}, \bibinfo{author}{P.~Castells}, \bibinfo{author}{P.~Gupta}, \bibinfo{author}{X.~Xu}, \bibinfo{author}{V.~Salaka},
\newblock \bibinfo{title}{Addressing cold start in product search via empirical bayes},
\newblock in: \bibinfo{booktitle}{CIKM}, \bibinfo{year}{2022}.
\bibitem[{Hendriksen et~al.(2020)Hendriksen, Kuiper, Nauts, Schelter, and de~Rijke}]{Hendriksen2020}
\bibinfo{author}{M.~Hendriksen}, \bibinfo{author}{E.~Kuiper}, \bibinfo{author}{P.~Nauts}, \bibinfo{author}{S.~Schelter}, \bibinfo{author}{M.~de~Rijke},
\newblock \bibinfo{title}{Analyzing and predicting purchase intent in e-commerce: Anonymous vs. identified customers},
\newblock in: \bibinfo{booktitle}{SIGIR eCom}, \bibinfo{year}{2020}.
\bibitem[{Rocchio(1971)}]{rocchio1971relevance}
\bibinfo{author}{J.~J. Rocchio},
\newblock \bibinfo{title}{Relevance feedback in information retrieval},
\newblock \bibinfo{journal}{The SMART retrieval system: experiments in automatic document processing}  (\bibinfo{year}{1971}).
\bibitem[{Santu et~al.(2017)Santu, Kanti, Parikshit, and Zhai}]{karmaker2017application}
\bibinfo{author}{K.~Santu}, \bibinfo{author}{S.~Kanti}, \bibinfo{author}{S.~Parikshit}, \bibinfo{author}{C.~Zhai},
\newblock \bibinfo{title}{On application of learning to rank for e-commerce search},
\newblock in: \bibinfo{booktitle}{SIGIR}, \bibinfo{year}{2017}, pp. \bibinfo{pages}{475--484}.
\bibitem[{Chen and Guestrin(2016)}]{chen2016xgboost}
\bibinfo{author}{T.~Chen}, \bibinfo{author}{C.~Guestrin},
\newblock \bibinfo{title}{Xgboost: A scalable tree boosting system},
\newblock in: \bibinfo{booktitle}{KDD}, \bibinfo{year}{2016}, pp. \bibinfo{pages}{785--794}.
\bibitem[{Liu(2009)}]{liu2009}
\bibinfo{author}{T.~Y. Liu},
\newblock \bibinfo{title}{Learning to rank for information retrieval},
\newblock \bibinfo{journal}{Foundations and Trends{\textregistered} in Information Retrieval} \bibinfo{volume}{3} (\bibinfo{year}{2009}) \bibinfo{pages}{225--331}.
\bibitem[{Chapelle et~al.(2012)Chapelle, Joachims, Radlinski, and Yue}]{chapelle2012large}
\bibinfo{author}{O.~Chapelle}, \bibinfo{author}{T.~Joachims}, \bibinfo{author}{F.~Radlinski}, \bibinfo{author}{Y.~Yue},
\newblock \bibinfo{title}{Large-scale validation and analysis of interleaved search evaluation},
\newblock \bibinfo{journal}{ACM Transactions on Information Systems} \bibinfo{volume}{30} (\bibinfo{year}{2012}) \bibinfo{pages}{1--41}.
\bibitem[{Bennett et~al.(2012)Bennett, White, Chu, Dumais, Bailey, Borisyuk, and Cui}]{Bennett2012}
\bibinfo{author}{P.~Bennett}, \bibinfo{author}{R.~White}, \bibinfo{author}{W.~Chu}, \bibinfo{author}{S.~Dumais}, \bibinfo{author}{P.~Bailey}, \bibinfo{author}{F.~Borisyuk}, \bibinfo{author}{X.~Cui},
\newblock \bibinfo{title}{Modeling the impact of short- and long-term behavior on search personalization},
\newblock in: \bibinfo{booktitle}{SIGIR}, \bibinfo{year}{2012}, pp. \bibinfo{pages}{185--194}.

\end{thebibliography}

\end{document}